# Citation Analysis with Microsoft Academic

Sven E. Hug[1,2,*], Michael Ochsner[1,3], and Martin P. Brändle[4,5]

[1] Social Psychology and Research on Higher Education, ETH Zurich, D-GESS, Muehlegasse 21, 8001 Zurich, Switzerland
[2] Evaluation Office, University of Zurich, 8001 Zurich, Switzerland
[3] FORS, 1015 Lausanne, Switzerland
[4] Zentrale Informatik, University of Zurich, 8006 Zurich, Switzerland
[5] Main Library, University of Zurich, 8057 Zurich, Switzerland
* Corresponding author. Tel.: +41 44 632 46 85, Fax: +41 44 634 43 79, Email: sven.hug@gess.ethz.ch

**Abstract**: We explore if and how Microsoft Academic (MA) could be used for bibliometric analyses. First, we examine the Academic Knowledge API (AK API), an interface to access MA data, and compare it to Google Scholar (GS). Second, we perform a comparative citation analysis of researchers by normalizing data from MA and Scopus. We find that MA offers structured and rich metadata, which facilitates data retrieval, handling and processing. In addition, the AK API allows retrieving frequency distributions of citations. We consider these features to be a major advantage of MA over GS. However, we identify four main limitations regarding the available metadata. First, MA does not provide the document type of a publication. Second, the "fields of study" are dynamic, too specific and field hierarchies are incoherent. Third, some publications are assigned to incorrect years. Fourth, the metadata of some publications did not include all authors. Nevertheless, we show that an average-based indicator (i.e. the journal normalized citation score; JNCS) as well as a distribution-based indicator (i.e. percentile rank classes; PR classes) can be calculated with relative ease using MA. Hence, normalization of citation counts is feasible with MA. The citation analyses in MA and Scopus yield uniform results. The JNCS and the PR classes are similar in both databases, and, as a consequence, the evaluation of the researchers' publication impact is congruent in MA and Scopus. Given the fast development in the last year, we postulate that MA has the potential to be used for full-fledged bibliometric analyses.

# Introduction

Microsoft Academic (MA) is a new service offered by Microsoft since 2015 and was introduced to the bibliometric research community by Harzing (2016). She assessed the coverage of this new tool by comparing the publication and citation record of her own oeuvre in Web of Science (WoS), Scopus, Google Scholar (GS), and MA. The Publish or Perish software (Harzing, 2007) was used to collect data from MA. Harzing (2016, p. 1646) finds that of the four competing databases "only Google Scholar outperforms Microsoft Academic in terms of both publications and citations" and concludes that MA is, with some reservations regarding metadata quality, an "excellent alternative for citation analysis" (p. 1647). She also conducted a citation analysis and calculated both the h-index and the hIa (Harzing, Alakangas, & Adams, 2014) for her oeuvre yet did not explore if other bibliometric analyses are feasible with MA. Hence, in this paper, we will explore if and how MA could be used for further bibliometric analyses. We will focus on Microsoft's Academic Knowledge API (AK API), an interface to access MA data. First, we will describe advantages and limitations of the AK API from the perspective of bibliometrics and compare it to GS, the closest competitor of MA. Second, we perform a citation analysis of researchers by normalizing data from MA and compare the results to those obtained with Scopus, an established database for bibliometrics.

# Academic Knowledge API

The AK API enables users to retrieve information from Microsoft Academic Graph (MAG). MAG is a database that models "the real-life academic communication activities as a heterogeneous graph consisting of six types of entities" (Sinha et al., 2015, p. 244). These entities are paper, field of study, author, institution (affiliation of author), venue (journal or conference series), and event (conference instances). Each of these entities is specified by entity attributes, which will be discussed below. Data for MAG is primarily collected from metadata feeds from publishers and web pages indexed by Bing (Sinha et al., 2015). MAG has grown massively from 2015 to 2016 and, according to Wade, Kuasan, Yizhou, and Gulli (2016), it contains approximately 140 million publication records (83)[1], 40 million authors (20), 3.5 million institutions (0.77), 60,000 journals (22,000), and 55,000 fields of study (50,000). Ribas, Ueda, Santos, Ribeiro-Neto, and Ziviani (2016) found that 59% of the papers in MAG are without citation information. Currently, MAG data can be accessed in three different ways: by using the MA search engine[2], by downloading historical snapshots of MAG[3], or by employing the AK API[4].

---

[1] Figures for 2015 are drawn from Sinha et al. (2015) and indicated in brackets.

The AK API offers the *Interpret*, the *Evaluate* and the *CalcHistogram* method for retrieving data from MAG. The latter two are essential for bibliometricians. The *Evaluate* method retrieves a set of attributes based on a query expression. Query expressions can be built with entity attributes (see below). An Evaluate request yields one or several matching results, or none, in case there is no match. Each result contains a natural log probability value to indicate the quality of the match. Thus, the *Evaluate* method is a means for collecting raw metadata from MA. In contrast, the *CalcHistogram* method calculates the distribution of attribute values for a set of paper entities. For example, it retrieves the distribution of the citations a journal has received in one year. Based on our exploration of the *CalcHistogram* method, it seems that the method can analyze around 2.4 million entities in one request. In order to calculate bibliometric indicators, however, data needs to be further processed.

In the AK API, there are 18 entity attributes that can be used to build query expressions as well as to specify the response of a query. Eight attributes are linked to the entity *paper*, four to the entity *author*, and two to each of the entities *field of study*, *journal*, and *venue* (entities in italics): *paper* – title, ID, year of publication, date of publication, citation count, estimated citation count, reference ID, words from title or abstract; *author* – name, ID, affiliation, affiliation ID; *field of study* / *journal* / *venue* – name, ID. In addition, there are 12 extended metadata attributes, which – in contrast to the 18 entity attributes – can only be used for specifying the query response. The 12 extended metadata attributes are available for the entities *paper* (ten attributes) and *venue* (two attributes): *paper* – volume, issue, first page, last page, DOI, display name of the paper, description (e.g. abstract), list of web sources of the paper, source format (e.g. HTML, PDF, PPT), source URL; *venue* – display name, short name. Based on our exploration of the AK API, it seems that almost all attributes that contain text are normalized. For example, the title of Immanuel Kant's *The Conflict of the Faculties* is stored in a normalized version (i.e. "der streit der fakultaten") of the original, non-normalized one ("Der Streit der Fakultäten"). However, there are some attributes in the AK API that do not seem to be normalized (i.e. display name of the paper, description of the paper, display name of the venue).

---

[2] https://academic.microsoft.com
[3] https://academicgraph.blob.core.windows.net/graph/index.html
[4] https://www.microsoft.com/cognitive-services/en-us/academic-knowledge-api

When comparing the six entities and 30 attributes available in the AK API with the metadata provided by GS (i.e. item ID, authors, title, source, year, volume, issue, pages, publisher, number of citations), it is obvious that metadata in MA is more structured than in GS and also considerably richer. Most importantly – and in contrast to GS –, MA-internal IDs are available for all entities as well as for the references of a paper. This will significantly facilitate data retrieval, handling and processing, and is a main advantage of MA over GS. As the studies of Prins, Costas, van Leeuwen, and Wouters (2016) and Bornmann, Thor, Marx, and Schier (2016) showed, data retrieval and handling with GS is extremely laborious due to metadata scarcity. We assume that the structure and richness of MA metadata will not only facilitate data handling but also translate into a wide variety of bibliometric indicators that can be calculated with MA. Wouters and Costas (2012) pointed out that GS provides very limited opportunities for calculating normalized indicators. As the empirical studies of Prins et al. (2016) and Bornmann et al. (2016) demonstrated, it is indeed possible to calculate normalized indicators with GS, but the process requires considerable effort and results are rather unsatisfactory. In contrast, we will show below that normalized indicators can be obtained with relative ease with MA. In comparison to WoS and Scopus, however, MA is substantially less equipped with regard to structure and richness of metadata. For example, in WoS, an author's reprint address alone comprises 31 attributes.[5] In conclusion, we think that the AK API has – due to the structure and richness of its metadata – the potential to be used for full-fledged bibliometric analyses (e.g. field-normalization, co-citation, bibliographic coupling, co-authorship relations, co-occurrence of terms).

A look at the attributes reveals not only strengths of MA but also weaknesses. First, there is no attribute for the type of the document. Without document type, distinguishing between citable and non-citable items will be very arduous if not impossible. Also, normalization of citation counts based on document type will prove to be difficult. Second, the DOI attribute cannot be used to build API requests, which would be beneficial for precision and sensitivity of the retrieval. Third, although MA has integrated a field attribute ("field of study"), it is unlikely that it can be deployed for field-normalization like the Subject Categories in WoS or the subject areas in Scopus. There are several reasons for this: The number of fields of study is growing as fields are created and updated by algorithms that exploit the keywords of papers (Sinha et al., 2015). As a consequence, there are currently more than 53,800 fields documented.[6] Fields are organized in four levels.[7] The highest level (L0) consists of the following 18 fields (in alphabetical order): art, biology, business, chemistry, computer science, economics, engineering, environmental science, geography, geology, history, materials science, mathematics, philosophy, physics, political science, psychology, and sociology. The second highest level (L1) includes for instance "social sciences", which is canonically considered to be a superordinate to L0 terms such as "psychology" or "sociology". In addition, on L1, there are fine-grained fields such as "Insurance score", "Titanic prime", and "Sonata cycle". Hence, field-normalization in MA likely has to be worked out without relying on the field attribute – or a meaningful hierarchy of fields has to be created. Similarly, in a longitudinal analysis of research topics, De Domenico, Omodei, and Arenas (2016) state – without giving reasons – that field information in MA is not suitable for classifying papers into disciplines.

---

[5] A description of WoS entities and attributes is available at
http://iuni.iu.edu/files/WoS_Documents/Entity_Relationship_Diagram_wos_core.pdf.

## Citation Analysis

In the next two sections, we will show how a comparative citation analysis of three researchers can be performed with the AK API. Since this paper focuses on feasibility and not on coverage and data quality, we will largely ignore the latter two topics in our analysis. As fields do not seem to be suitable reference sets in MA, we follow the steps of Bornmann et al. (2016), who selected journal and year as benchmarking units in their GS evaluation exercise. We calculated the journal normalized citation score (JNCS) as outlined by Rehn, Wadskog, Gornitzki, and Larsson (2014), which belongs to the family of average-based indicators, such as the MNCS (Waltman, van Eck, van Leeuwen, Visser, & van Raan, 2011). In addition, we calculated percentile rank (PR) classes (Bornmann, Leydesdorff, & Mutz, 2013), which belong to the family of distribution-based indicators. To test if meaningful results can be obtained with MA, we compare MA values with those obtained with Scopus, an established database for bibliometrics.

**Data Collection and Analysis**

The journal *Scientometrics* constitutes the reference set of our analysis. We selected three researchers, who contributed comparable numbers of publications to Scientometrics from 2010 to 2014, and searched their publications in the journal (n=57). Based on the titles of these publications, we then extracted metadata (including citation counts) from MA using the *Evaluate* method of the AK API. All publications were found in MA. Based on the authors' names, we checked if additional publications were listed in MA, which was not the case. Similarly, we extracted metadata from Scopus. All publications were found in Scopus and no additional ones were identified. However, in MA, we encountered issues regarding the quality of the metadata. 11 publications had a wrong publication year (plus or minus one year). Much more severely, we found that one of the three authors is not listed as an author on 64% of his publications. All data was collected in the first week of September 2016.

---
[6] https://academicgraphwe.blob.core.windows.net/graph-2016-02-05/FieldsOfStudy.zip
[7] https://academicgraphwe.blob.core.windows.net/graph-2016-02-05/FieldOfStudyHierarchy.zip

Since the document type is missing in the metadata of MA, we included all publications in *Scientometrics* from 2010 to 2014 to build the reference set. Based on the journal ID of *Scientometrics* in MA, we retrieved the citation distribution for every year by using the *CalcHistogram* method. The query yielded a total of 1,300 publications and 11,485 citations. Applying the same search logic to Scopus, we collected the yearly citation distributions of *Scientometrics* based on its ISSN. The Scopus search yielded slightly more publications (1,392) as well as citations (12,954). We did not check the overlap of the two reference sets.

The journal normalized citation score, JNCS, was calculated as follows: "The number of citations of [each author's] publications is normalized by dividing it with the world average of citations to publications [...] published the same year in the same journal. The indicator is the mean value of all the normalized citation counts for the [author's] publications" (Rehn et al., 2014, p. 14). If the calculated value is greater (or smaller) than 1.0, this means that the author's publications are cited more (or less) frequently than the average of the publications in the journal. We calculated PR classes following the procedure outlined by Bornmann et al. (2013). We sorted the publications of the reference set in descending order by their number of citations and assigned publications with 0 citations a percentile of 0 and calculated the remaining percentiles from the citation distribution of the reference set. We then assigned each of the authors' publications to one of four PR classes. PR class 4 consists of publications with a percentile equal to or larger than the $90^{th}$ percentile (i.e. the top 10% most cited publications), PR class 3 of publications with a percentile equal to or larger than the $80^{th}$ percentile and smaller than the $90^{th}$ percentile, PR class 2 of publications with a percentile equal to or larger than the $50^{th}$ percentile and smaller than the $80^{th}$ percentile, and PR class 1 of publications with a percentile smaller than the $50^{th}$ percentile (i.e. the 50% least cited publications). Since distributions of citations are discrete and publications often have the same number of citations, it is usually difficult to define the threshold of PR classes without introducing biases (Waltman & Schreiber, 2013). One way to deal with this issue is to choose thresholds according to the citation distribution at hand. Since the 20$^{th}$ percentile fits our data, we use the 20$^{th}$ percentile as a threshold for PR class 3 and not the 25$^{th}$ percentile, which is often used as a threshold (see Bornmann et al., 2013), but does not fit our data. As we noted above, not all publications in MA were assigned to the correct publication year. In order not to distort the data, we calculated both the JNCS and the PR classes for MA with the publication years assigned in MA, and for Scopus with the publication years assigned in Scopus.

**Results**

The JNCS in MA and Scopus is 1.30 and 1.42 for researcher A, 0.65 and 0.58 for researcher B, and 0.55 and 0.69 for researcher C, respectively. Hence, the values differ slightly between MA and Scopus. Moreover, researchers B and C swap places if the three researchers are ranked according to their JNCS in MA and Scopus (see Table 1). Nevertheless, the overall assessment in MA and Scopus stays the same. While the publication impact of researcher A is clearly above the journal's average, the impacts of researchers B and C are clearly below it.

Table 1    Journal normalized citation score (JNCS) of researchers' publications

|  | Researcher A | | Researcher B | | Researcher C | |
|  | JNCS | Rank[1] | JNCS | Rank | JNCS | Rank |
| --- | --- | --- | --- | --- | --- | --- |
| Microsoft Academic | 1.30 | 1 | 0.65 | 2 | 0.55 | 3 |
| Scopus | 1.42 | 1 | 0.58 | 3 | 0.69 | 2 |

*Note:* Rank = rank of researcher according to her / his JNCS.

Table 2  Publications of researchers in percentile rank classes

|  | PR class[1] | Percentile interval | Researcher A Per cent[2] | Researcher B Per cent | Researcher C Per cent |
|---|---|---|---|---|---|
| Microsoft Academic | 4 | [90th; 100th] | 28 | 0 | 0 |
|  | 3 | [80th; 90th[ | 12 | 29 | 11 |
|  | 2 | [50th; 80th[ | 20 | 21 | 17 |
|  | 1 | [0th; 50th[ | 40 | 50 | 72 |
| Scopus | 4 | [90th; 100th] | 20 | 0 | 0 |
|  | 3 | [80th; 90th[ | 16 | 21 | 11 |
|  | 2 | [50th; 80th[ | 32 | 36 | 28 |
|  | 1 | [0th; 50th[ | 32 | 43 | 61 |

*Note:* 1 = Percentile rank class; PR class 4 is the class with the highest impact (i.e. it comprises the top 10% most cited publications); 2 = Percentage of an authors' publications in a PR class.

The shares of each authors' publications in the four PR classes are given in Table 2. The distributions of the researcher's publications in the four PR classes do not differ considerably between MA and Scopus. Hence, the performance of the researchers is assessed similarly in MA and Scopus. If we employ the top 10% percentiles (i.e. PR class 4) to tag high performing publications, which is often done in evaluative bibliometrics (Tijssen, Visser, & van Leeuwen, 2002; Waltman & Schreiber, 2013), we can conclude that both MA and Scopus indicate a high performance for researcher A but not for researchers B and C.

**Conclusion**

We explored if and how MA could be used for bibliometric analyses. First, we examined the AK API, an interface to access MA data. Second, we performed a citation analysis of three researchers by normalizing data from MA and compared the results to those obtained with Scopus. The AK API enables users to retrieve information from MA. We highlighted that MA has grown massively from 83 million publication records in 2015 to 140 million in 2016. We described how users could retrieve raw metadata as well as calculate frequency distributions of citations with the AK API. These two functions are not available for GS. We found that the metadata in MA is clearly more structured than in GS, which the article of Harzing (2016) has already implied, and it is also considerably richer. Most importantly, MA-internal IDs are available for papers, references, authors, affiliations, fields of study, journals and venues (i.e.

journal or conference series). This significantly facilitates data retrieval, handling and processing and is a major advantage of MA over GS. As the studies of Prins et al. (2016) and Bornmann et al. (2016) showed, data retrieval and handling as well as creating normalized indicators with GS is extremely laborious and rather unsatisfactory. In contrast, we retrieved and handled data from MA without much effort and obtained an average-based indicator (i.e. the JNCS) as well as a distribution-based indicator (i.e. PR classes) with relative ease. Hence, MA has an edge over GS with respect to calculating indicators and therefore is more suitable for evaluative bibliometrics. We postulate that MA has – based on these features – the potential to be used for full-fledged bibliometric analyses (e.g. field-normalization, co-citation, bibliographic coupling, co-authorship relations, co-occurrence of terms). However, our exploration of MA reveals four main limitations regarding the available metadata. First, MA does not provide the document type of a publication. Second, the "fields of study" are dynamic, too specific and field hierarchies are incoherent. Hence, normalization in MA likely has to be worked out without relying on the field attribute and the document type. Third, some publications are assigned to incorrect years, an issue that Harzing (2016) has already highlighted. In particular, we found that 19% of the publications had an incorrect publication year (plus or minus one year). Fourth, the metadata of some publications did not include all authors. In particular, we found that one of the analyzed authors is not listed as an author on 64% of his publications. This brings to mind the authorship parsing problems of GS put forward by Jacso (2010) some years ago. However, in our case, the author was just omitted and not replaced by a "phantom author". Since the third and fourth limitation is based on a small sample size, future studies are needed in order to assess these issues on a larger scale. Furthermore, there is another minor but not severe limitation of the AK API, namely that the DOI of a paper cannot be used to build API requests even though it is stored in MA and can be retrieved. Integration of the DOI in the query expression would be beneficial for precision and sensitivity of the retrieval.

We showed that average-based indicators as well as distribution-based indicators can be calculated with MA and that normalization of citation counts is therefore feasible with MA. We found that the JNCS of three researchers differ marginally between MA and Scopus and that the evaluation of the publication impact is hence congruent in both databases. While the impact of researcher A is clearly above the average of the reference set, the impacts of researchers B and C are clearly below it. Similarly, the distribution of researchers' publications in PR classes did differ only slightly between MA and Scopus and, hence, the publication impact of the three researchers is assessed congruently in the two databases. When focusing on those publications that rank in PR class 4 (i.e. the publications which belong to the top 10% most frequently cited of the reference set), we found that – both in MA and Scopus – researcher A has a high impact in contrast to researchers B and C. These results are in line with Harzing (2016) and Harzing & Alakangas (2016)[8] who found that both the h-index and the hIa were similar in MA and Scopus. Hence, citation analyses with MA and Scopus seem to yield uniform results.

In her study on the coverage of MA, Harzing (2016, p. 1646) concludes that "only Google Scholar outperforms Microsoft Academic in terms of both publications and citations". Based on our exploration of MA, we conclude that MA outperforms GS in terms of functionality, structure and richness of data as well as with regard to data retrieval and handling. Our conclusions are, however, highly dependent on coverage issues and metadata quality, which were not the focus of this paper. Therefore, further studies are needed to assess the suitability of MA as a bibliometric tool. Nevertheless, we hope that MA cannot only "trigger a new horizon of research efforts towards defining new academic impact metrics", as Microsoft expressed it (see Sinha et al., 2015, p. 243), but also become a useful tool for calculating established bibliometric indicators.

---

[8] Data collection and publication of Harzing & Alakangas' (2016) study took place after the submission of this paper.